\documentclass[aps,pra,superscriptaddress,twocolumn,twocolumn,10pt]{revtex4-1}
\usepackage{amsmath}
\usepackage{latexsym}
\usepackage{graphicx}
\usepackage{amsfonts}
\usepackage{braket}
\usepackage{xcolor}
\definecolor{darkblue}{rgb}{0,0,0.5} 
\usepackage{transparent}
\usepackage[colorlinks=true,
        urlcolor=darkblue,
        anchorcolor=darkblue,
        linkcolor=darkblue,
        citecolor=darkblue,
        pdfkeywords={SVG; LaTeX; Inkscape}]{hyperref}
\usepackage{natbib}
\usepackage{amssymb}
\usepackage{nicefrac}
\usepackage{fancyhdr}
\usepackage[utf8]{inputenc}
\usepackage{dsfont}
\usepackage{colortbl}
\usepackage{xcolor}
\usepackage{subfigure}
\usepackage{psfrag}
\usepackage{tikz}
\usepackage{paralist}
\usepackage{nicefrac}
\usepackage{ulem}
\usepackage[section]{placeins}
\usepackage{mathtools}
\usepackage{verbatim}
\usepackage{balance}
\usepackage{lastpage}
\mathtoolsset{showonlyrefs}

\newcommand{\pd}[2]{\frac{\partial #1}{\partial #2}}
\newcommand{\spd}[3]{\ifthenelse{\equal{#2}{#3}}{\frac{\partial^2 #1}{\partial {#2}^2}}{\frac{\partial^2 #1}{\partial #2 \partial #3}}}
\newcommand{\im}{\mathrm{i}}

\newcommand{\ahorizon}{a sonic horizon}
\newcommand{\thehorizon}{the sonic horizon}

\usetikzlibrary{arrows,shapes,decorations.pathmorphing}

\makeindex

\begin{document}

\title{Quantum simulation of particle creation in curved space-time}

\author{R. P. Schmit}
\affiliation{Theoretical Physics, Saarland University, Campus, 66123 Saarbr{\"u}cken, Germany}
\author{B. G. Taketani}
\affiliation{Departamento de F\'isica, Universidade Federal de Santa Catarina, 88040-900, Florian\'opolis, SC, Brazil}
\author{F. K. Wilhelm}
\affiliation{Theoretical Physics, Saarland University, Campus, 66123
Saarbr{\"u}cken, Germany}

\begin{abstract}
Conversion of vacuum fluctuations into real particles was first predicted by L. Parker considering an expanding universe, followed in S. Hawking's work on black hole radiation. Since their experimental observation is challenging, analogue systems have gained attention in the verification of this concept. Here we propose an experimental set-up consisting of two adjacent piezoelectric semiconducting layers, one of them carrying dynamic quantum dots (DQDs), and the other being p-doped with an attached gate on top, which introduces a space-dependent layer conductivity. The propagation of surface acoustic waves (SAWs) on the latter layer is governed by a wave equation with an effective metric. In the frame of the DQDs, this space- and time-dependent metric possesses \ahorizon~for SAWs and resembles that of a two dimensional non-rotating and uncharged black hole to some extent. The non-thermal steady state of the DQD spin indicates particle creation in form of piezophonons.
\end{abstract}

\maketitle
%%%%%%%%%%%%%%%%%%%%%%%%%%%%%%%%%%%%%
%%%%%%%%%%%%%%%%%%%%%%%%%%%%%%%%%%%%%
\section{Introduction}
The ubiquitous presence of vacuum fluctuations is arguably one of the most surprising effects of quantum theory. Their existence is indirectly observable via the modification of the electron's magnetic moment~\cite{item2} or the Lamb shift of an atomic spectrum~\cite{item1}. A more direct access could be accomplished by converting the virtual particles into directly observable real ones. Such particle creation is predicted to take place under various conditions such as the dynamical Casimir effect~\cite{yablonovitch1989accelerating, schwinger1992casimir} and related circumstances~\cite{bruschi2012voyage}, during the expansion of the universe~\cite{parker1968particle}, or due to the presence of a black hole's event horizon~\cite{hawking1975particle}. While the dynamical Casimir effect has been experimentally verified~\cite{wilson2011observation}, direct experimental investigations for the two latter theoretical predictions are more challenging. Black hole analogues were first considered by W. Unruh~\cite{unruh1981experimental} as a means to overcome this difficulty. He discovered that the propagation of sound waves in an irrotational fluid is governed by an effective metric matching that of a gravitating spherical, non-rotating massive body in Painlev\'e-Gullstrand coordinates~\cite{painleve1922mecanique} with line element
\begin{align}
 \mathrm{d}s^2 = -\left[ c_s^2 - v^2(r) \right]\mathrm{d}t^2 + 2v(r)\mathrm{d}r\mathrm{d}t + \mathrm{d}r^2 + r^2\mathrm{d}\Omega^2 \label{PG_metric}
\end{align}
with the speed of sound $c_s$ and flow speed $v(r)$ of the fluid. Similarly, physical systems~\cite{fedichev2003gibbons, fischer2004quantum, eckel2018rapidly, fey2018ion, wittemer2019particle} giving rise to an effective metric matching the Friedmann-Lema\^{i}tre-Robertson-Walker line element
\begin{align}
 \mathrm{d}s^2 = -c^2 \mathrm{d}t^2 + a^2(t)\mathrm{d}\textbf{r}^2 \label{FLRW_metric}
\end{align}
can mimic a universe that expands according to the scale factor $a(t)$.

A vast number of analogue systems have been investigated, such as black hole analogues in liquid Helium~\cite{jacobson1998event}, dc-SQUID transmission lines~\cite{nation2009analogue}, electromagnetic wave-guides~\cite{schutzhold2005hawking}, water waves~\cite{rousseaux2008observation,Weinfurtner2011,euve2016}, hydrodynamic microcavity polariton flow~\cite{nguyen2015acoustic}, optical set-ups~\cite{elazar2012all,philbin2008fiber,philbin2008fiber,marino2008acoustic,belgiorno2010hawking} and Bose-Einstein condensates~\cite{Steinhauer2016observation,cropp2016analogue}. Systems analogue to particle creation in the dynamical Casimir effect~\cite{lange2018,vezzoli2019} and in an expanding universe~\cite{schuetzhold2007,fey2018ion,eckel2018rapidly} have also been proposed in systems ranging from trapped ions and BEC to photonic crystal fibers.

\begin{figure}
 \centering
 \includegraphics{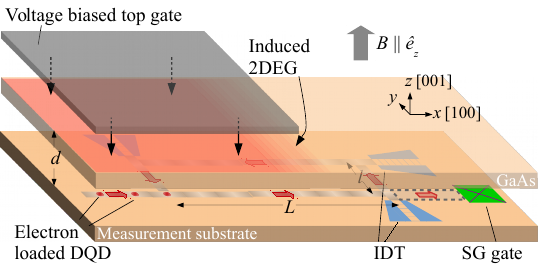}
 \caption{{\bf Partial sketch of the experimental set-up.} The formation of \thehorizon~for surface acoustic waves on the upper GaAs layer is due to the inhomogeneous two-dimensional electron gas (2DEG) induced by the attached gate. The electrons in the dynamic quantum dots (DQDs) on the lower GaAs layer serve as detector for the created particles in form of piezoelectric phonons: The thermal occupation of the two electron-spin states in the DQDs is expected to be altered due to their spin-orbit interaction with the created phonons. A Stern-Gerlach (SG) gate allows for the readout of the electronic spin. The arrangement of the interdigitated transducers (IDTs) serves as a storage ring for the DQDs in order to provide enough time for interaction. Not shown is the cryogenics.}
 \label{fig:setup}
\end{figure}

In this paper, we present an experimental set-up (see Fig~\ref{fig:setup}) with features resembling both analogues of a black hole~\cite{jacobson1998event, nation2009analogue,schutzhold2005hawking,rousseaux2008observation, Weinfurtner2011, nguyen2015acoustic,elazar2012all,philbin2008fiber,marino2008acoustic,belgiorno2010hawking,Steinhauer2016observation} and of an expanding universe~\cite{fedichev2003gibbons, fischer2004quantum, eckel2018rapidly, fey2018ion, wittemer2019particle} and discuss particle creation in this system. We investigate the 1-dimensional propagation of surface acoustic waves (SAWs) on a piezoelectric p-type semiconducting substrate with an inhomogeneous 2-dimensional electron gas (2DEG)~\cite{simon1996coupling,Rahman2006,Shao2015}. These systems have been widely used, e.g. to probe quantum effects in 2DEGs~\cite{Efros1990}, as well as for electron transport~\cite{McNeil2011,Hermelin2011}. SAWs have also been proposed as a quantum computation platform~\cite{barnes2000quantum} and for quantum simulations~\cite{Gustafsson2014}. An attached gate allows for a controlled spatial modulation of the 2DEG density~\cite{sze2008semiconductor}. Similar to Unruh's proposal~\cite{unruh1981experimental}, this introduces a space-dependent modulation of the SAW speed. For an observer moving along the SAWs the substrate corresponds to a flowing medium carrying excitations. In the rest frame of the moving observer the SAW propagation will be described by an effective metric giving rise to \ahorizon~for SAWs at the crossover between subsonic and supersonic propagation. This indicates particle creation in form of piezoelectric phonons. A possible implementation of a moving particle detector using dynamic quantum dots (DQDs) is also discussed.

This paper is organized as follows. In Sec~\ref{sec:blocks} we demonstrate the existence of \ahorizon~present for a moving observer due to a modulation of the speed of sound in a piezoelectric substrate, and describe how such a modulation can be achieved using a biased gate. In Sec~\ref{sec:res} we describe a measurement set-up using dynamic quantum dots and argue that they equilibrate to a non-thermal state when in contact with a zero-temperature phonon bath. This is attributed to particle creation in form of piezoelectric phonons. In Sec~\ref{sec:conc} we present concluding remarks.

\section{Building blocks}
\label{sec:blocks}
In this section, we will describe the fundamental requirements to generate \ahorizon~for SAWs. In adapting the original proposal of Ref~\cite{unruh1981experimental} to solid-state devices, the difficulty lies in designing a moving medium for wave propagation. This can be circumvented by having a space- (and time-) dependent wave speed $c(x, t)$. An observer moving with a proper speed $v$ along the waves will eventually experience the crossover between subsonic and supersonic propagation - a direct route for the formation of \ahorizon. Consequently, this opens up the necessity of a moving detector for measuring the created phonons. Here, the local modulation of the SAW speed can be achieved by exploiting its dependence on the substrate conductivity~\cite{simon1996coupling, ingebrigtsen1970linear, gumbs1998interaction, bierbaum1972interaction, hutson1962elastic, wixforth1989surface}, which can be changed locally by biasing a thin gate attached to the piezoelectric p-type semiconducting substrate (see, e.g. Ref~\cite{sze2008semiconductor}): Biasing the gate with a voltage induces a 2DEG in the surface of the substrate in the vicinity to the gate, changing the substrate conductivity. A properly moving detector experiences \ahorizon~for SAWs. The detector comprises of dynamic quantum dots (DQDs) transporting photogenerated electrons and propagating in an adjacent substrate (labeled as measurement substrate in Fig~\ref{fig:setup}). A magnetic field along the $[001]$ axis leads to a Zeeman-splitting of the electron-spin states. Due to spin-orbit interaction between the electrons and the created piezoelectric phonons, the thermal occupation among the spin states is expected to be altered and can be read out by a Stern-Gerlach (SG) gate~\cite{barnes2000quantum, furuta2004single}, which converts spin into current paths via the Stern-Gerlach effect. The arrangement of interdigitated transducers (IDTs) serves as storage ring for the electrons and enables for a sufficiently long interaction time required for equilibration (estimated to $\sim$ 1 s, see Sec~\ref{sec:res}).

We will now derive the effective metric and present set-up details.

\subsection{Effective metric}
\label{sec:metric}
We consider the 1-dimensional propagation of SAWs along the $x$-direction with a space-dependent speed of sound $c(x)$. With $u$ denoting the SAW amplitude, the dynamics follows the usual wave equation (see, e.g.~\cite{datta1986surface})~\cite{Note1}
\begin{align}
 \frac{\partial^2 u}{\partial^2 t} = \frac{\partial}{\partial x}\left( c^2(x) \frac{\partial u}{\partial x} \right) \label{eq:wave_eq}.
\end{align}
This equation describes wave propagation in a space-time with line element given by+
\begin{align}
 \mathrm{d}s^2 = -c^2(x)\mathrm{d}t^2 + \mathrm{d}x^2.
 \label{eq:static_metric}
\end{align}
The Galilean transformation to the reference frame of an observer moving at speed $v$ along the $x$-direction is accomplished by the substitutions 
\begin{align}
t \to t\quad,\quad x \to x + vt 
\end{align}
\begin{align}
 \frac{\partial}{\partial t} \to \frac{\partial}{\partial t} - v\frac{\partial}{\partial x}\quad,\quad \frac{\partial}{\partial x} \to \frac{\partial}{\partial x}
\end{align}
in Eq~\eqref{eq:wave_eq}. Note that $c$ is now a function of $x$ and $t$. The effective metric describing the SAW dynamics in the moving reference frame leads to the line element
\begin{align}
 \mathrm{d}s^2 = -\left[c^2(x-vt)-v^2\right] \mathrm{d}t^2 + 2 v \mathrm{d}t \mathrm{d}x + \mathrm{d}x^2,
 \label{eq:dynamic_metric}
\end{align}
revealing \ahorizon~where $c^2 - v^2 = 0$. Not that the wave equation and the associated effective metric are given in the rest frame of the substrate. In this coordinate system the effective metric is (only) space-dependent stemming from the space dependence of the speed of sound. The Galilean boost into the frame of the DQDs introduces the time-dependence in the new coordinate system.

This effective metric has features in common with Painlevé-Gullstrand's metric and the Friedmann-Lema\^{i}tre-Robertson-Walker metric revealing its potential for particle creation, which is supported by calculations using the Bloch-Redfield equations in Sec~\ref{sec:res}. However, we note that the present set-up does not strictly simulate either of both systems, as Eq~\eqref{eq:dynamic_metric} does not match their respective metric precisely.

\subsection{SAW dynamics}
\label{sec:saw_speed}
The propagation of SAWs can be controlled by their interaction with a 2DEG in the substrate~\cite{simon1996coupling, ingebrigtsen1970linear, gumbs1998interaction, bierbaum1972interaction, hutson1962elastic, wixforth1989surface}. Here we follow Ref~\cite{hutson1962elastic}, which presents a detailed calculation of SAW propagation on a piezoelectric semiconductor with a \textit{homogeneous} electron gas, and extend their results to the inhomogeneous case. \\
Due to the piezoelectric effect the SAW is accompanied by an electric field, thus interacting with the 2DEG and inducing currents that dissipate energy due to Ohmic losses. The piezoelectric effect is taken into account by introducing a space dependent charge density $n_s$ that obeys Maxwell's equation
\begin{align}
 \pd{D}{x} = -q n_s
\end{align}
in one dimension with the elementary charge denoted by $q = |q|$ ($e$ will be reserved for the piezoelectric constant) and where $D$ is the electric displacement field. The induced current density is given by
\begin{align}
 j(x, t) = -q\left[n_{\text{2DEG}}(x) + fn_s(x, t)\right]\mu E(x, t),
\end{align}
where $\mu$ denotes the electron mobility, $f$ accounts for the part of the induced space charge being in the conduction band (for a calculation of $f$, see~\cite{hutson1962elastic}), $n_{\text{2DEG}}$ denotes the \textit{time-independent} density of the 2DEG induced by an attached gate and $E(x,t)$ is the electric field. The time-independence of $n_{\text{2DEG}}$ is necessary in order to recover the usual wave equation for the SAW amplitude. Diffusion currents $\sim k_B T \frac{\partial}{\partial x}\left( n_{\text{2DEG}} + fn_s \right)$ due to spatial inhomogeneous charge distribution can be neglected in the low-temperature limit we propose to work in. The total charge density contributing to the electric current is given by
\begin{align}
 \rho = -q\left[ n_{\text{2DEG}} + fn_S \right].
\end{align}
Using the continuity equation for the charge current and density
\begin{align}
 \pd{\rho}{t} + \pd{j}{x} = 0
\end{align}
and the time-independence of $n_{\text{2DEG}}$, one can derive an equation relating $D$ and $E$,
\begin{align}
 - \frac{\partial^2 D}{\partial x \partial t} = \mu \frac{\partial}{\partial x} \left( \left[ f\frac{\partial D}{\partial x} - q n_{\text{2DEG}} \right]E \right).
\end{align}
Using a plane-wave ansatz
\begin{align}
 E &= E_0 \exp\left\{ \im \left( k(x) - \omega t\right) \right\} \\
 D &= D_0 \exp\left\{ \im \left( k(x) - \omega t\right) \right\},
\end{align}
and neglecting terms with the product $E \cdot D$~\cite{Note2}, one can write $D = \varepsilon_{\text{eff}}E$ with an effective permittivity
\begin{align}
 \varepsilon_{\text{eff}} = \frac{\mu q}{\omega} \left[ \frac{\partial_x n_{\text{2DEG}}}{\partial_x k} + \im n_{\text{2DEG}} \right], \label{eq:eff_perm}
\end{align}
where $\partial_x$ abbreviates the spatial derivative. The equations of state for a piezoelectric material 
\begin{align}
 T &= dS - eE \\
 D &= eS + \varepsilon E,
\end{align}
where $T$ and $S$ denote stress and strain constants, $d$ is the elastic constant and $e$ is the piezoelectric constant, can be simplified to $T = d_{\text{eff}}S$ with an effective elastic constant
\begin{align}
 d_{\text{eff}} = d\left[ 1 + \frac{e^2}{\varepsilon d} \left( 1- \frac{\varepsilon_{\text{eff}}}{\varepsilon} \right)^{-1} \right]. \label{eq:eff_d}
\end{align}
This equation illustrates the effect of piezoelectric stiffening, i.e. a dressed elastic constant due to the piezoelectric effect~\cite{johannsmann2015quartz}. The equation of motion for the SAW amplitude $u$ is given by
\begin{align}
 S = \frac{\partial u}{\partial x} \text{ and } \frac{\partial T}{\partial x} = \rho \frac{\partial^2 u}{\partial t^2},
\end{align}
leading, with Eq~\eqref{eq:eff_d}, finally to the wave equation of Eq~\eqref{eq:wave_eq} with the SAW speed given by
\begin{align}
 c(x) = \rm Re \left( \sqrt{\frac{d_{\text{eff}}}{\rho}} \right) \label{eq:SAW_speed_1}.
\end{align}
Note that the RHS of Eq~\eqref{eq:SAW_speed_1} also depends on the SAW speed via $\partial_xk(x) = \omega / c(x)$. Thus, solving Eq~\eqref{eq:SAW_speed_1} for $c(x)$ gives, in principle, an expression for the SAW speed in terms of the 2DEG density. An approximate solution is given in Appendix~\ref{app:SAW_speed}.

\subsection{2DEG density modulation}
\label{sec:density_modulation}
\begin{figure}
 \centering
 \def\svgwidth{\columnwidth}
 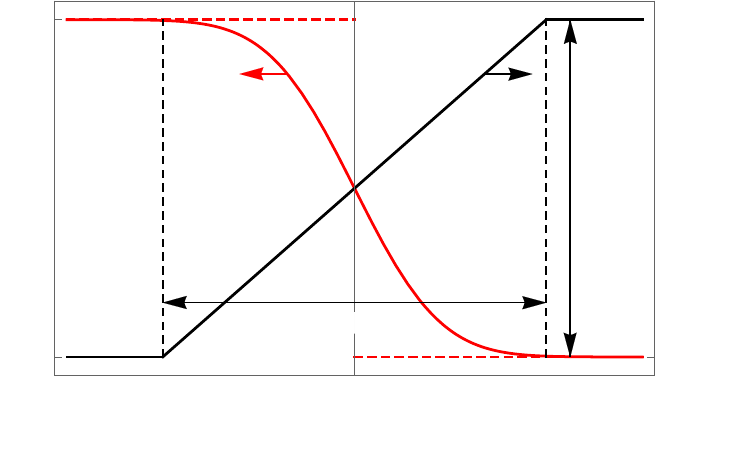
 \caption{{\bf Spatial profiles around the gate's edge.} Shown are the densities of the approximate (dashed red) and actual, smeared out (solid red) charge distribution of the induced 2DEG. The latter one arises from the first one due to screening effects inside the semiconducting substrate, which smoothen the charge density in a narrow region of approximate thickness $4 \kappa_s^{-1}$ around the gate's edge with the screening length $\kappa_s^{-1}$. The corresponding SAW speed $c(x)$ (black) according to Eq~\eqref{eq:SAW_speed_1} takes the values $c_0$ and $c_0(1+K^2/2)$ with the piezoelectric coupling constant $K^2$ in the region with high and low 2DEG density, respectively, and approximately aligns linearly in the transition region.}
 \label{fig:graph}
\end{figure}

In this section we describe the charge distribution in the 2DEG which is induced by the attached gate biased with a voltage $V_G \sim 10$ V. As a bare approximation the 2DEG is assumed to distribute homogeneously over the area of the gate,
\begin{align}
 n_0(x) = n_{\text{max}}H\!\left(-x \right)\; ,
 \label{eq:n_0}
\end{align}
with a density amplitude $n_{\text{max}}$ proportional to the applied gate voltage $V_G$~\cite{sze2008semiconductor}, and the Heaviside step function $H(x)$. However, the actual 2DEG is smeared out and its density $n_{\text{2DEG}}$ is smoothed close to the gate's edge (illustrated in Fig~\ref{fig:graph}). We take this into account by convoluting the approximated density $n_0(x)$ and a Gaussian with FWHM denoted by $\kappa_s^{-1}$, reading
\begin{align}
 G(x) &= A \exp\!\left[-(\kappa_s x)^2\right], \\
 n_{\text{2DEG}}(x) &= (G*n_0)(x) \\
 &= \frac{n_{\text{max}}}{2}\left[1-\text{erf}\left(\kappa_s x\right)\right], \label{eq:2DEG_dens}
\end{align}
where ``$*$'' denotes the convolution operation. The normalization coefficient $A$ guarantees charge conservation, 
\begin{align}
 \int \limits_{-\infty}^{\infty}\left[n_{\text{2DEG}}(x) - n_0(x)\right]\mathrm{d}x=0.
\end{align}

The phenomenological parameter $\kappa_s^{-1}$ should be of the order of the screening length of the substrate material which, for moderate doping of the p-type semiconducting substrate, is typically of the order of $10^{-8}$ m~\cite{mahan2013many, stern1974low}. This is the density that will modulate the speed of sound for the SAWs in Eq~\eqref{eq:SAW_speed_1}.

% Results and Discussion can be combined.
\section{Particle creation and detection}
\label{sec:res}
The comparison of the two effective metrics with line elements Eqs~\eqref{eq:dynamic_metric} and~\ref{eq:static_metric} shows that \ahorizon~is present only in the reference frame of the moving observer. Consequently, particle creation is expected to occur and to be measured only in the moving reference frame as well. This is a result of the observer dependent notion of particles (see, e.g.~\cite{fedichev2004observer})~\cite{note6}. For this purpose, we propose to use moving electrons trapped in the lateral piezoelectric potentials accompanying SAWs, also known as dynamic quantum dots (DQDs), as proposed in~\cite{rocke1997acoustically, stotz2005acoustically}. The DQDs move on the measurement substrate, so that their speed $v$ is constant and does not follow the speed profile $c(x)$ of the upper substrate (see Fig~\ref{fig:setup}). In this section we show that the DQDs reach a non-thermal steady-state, even when interacting with a zero-temperature phononic bath, which we argue is due to the particle creation in the moving frame.

We choose a coordinate system with axes $x$ and $z$ pointing along the [100] and [001] axes of GaAs. Applying a magnetic field $B \sim 1$ T~\cite{Note3} along the $z$ axis leads to Zeeman-splitting of the two electron-spin states denoted by $\ket{\uparrow}$ and $\ket{\downarrow}$, respectively, with an energy separation $E_{\uparrow} - E_{\downarrow} = E_0 = g \mu_B B$. The Dresselhaus spin-orbit coupling~\cite{dresselhaus1955spin} is the dominant process for spin-flip transitions between the Zeemann sublevels in the DQD~\cite{stotz2005acoustically, khaetskii2001spin}.

For a motion of the DQD in $x$ direction, $\textbf{v} = v \hat{e}_x$, the Hamiltonian approximately describing the electron-spin while interacting with the piezoelectric phonons can be written as $H = H_S + H_{\text{SB}} + H_B$ with the spin Hamiltonian~\cite{zhao2016doppler,huang2013spin,golovach2004phonon,borhani2006spin,golovach2006electric,huang2014electron}
\begin{align}
 H_S = E_0 \sigma_z + \Delta E \sigma_y,
\end{align}
the phonon bath Hamiltonian
\begin{align}
 H_B = \sum \limits_{q}\hbar \omega_q \left( b_q^{\dagger} b_q + \frac{1}{2}\right)
\end{align}
and their mutual interaction Hamiltonian
\begin{align}
 H_{\text{SB}} = \sigma_x \sum \limits_q M_q {\rm e}^{\im qvt} \left( b_{-q}^{\dagger}  + b_q\right).
\end{align}
Here, $\Delta E$ originates from a motion induced constant magnetic field, and the spin-phonon coupling is due to a motion induced magnetic noise originating from the electric noise due to the piezophonons~\cite{zhao2016doppler,huang2013spin}. Details about the parameters and their dependence on the DQD speed and the material constants can be found in Ref~\cite{zhao2016doppler} and references therein. Using this model Hamiltonian, the dynamics of the DQD spin while interacting with the phonon system can be calculated using the Bloch-Redfield equations, which read~\cite{weiss2012quantum}
\begin{align}
 \dot{\rho}_{\mu \nu}(t) = -\im \omega_{\mu \nu}\rho_{\mu \nu}(t) + \sum \limits_{\kappa \lambda}R_{\mu \nu \kappa \lambda} \rho_{\kappa \lambda}(t). \label{eq:BRE}
\end{align}
Here, $R_{\mu \nu \kappa \lambda}$ are the elements of the Redfield tensor and the $\rho_{\mu \nu}$ are the elements of the reduced density matrix of the DQD in the eigenstate basis of $H_S$ denoted by $\ket{0}$ and $\ket{1}$ for the ground and excited state, respectively. For eigenstates $\ket{\mu}$ and $\ket{\nu}$ of $H_S$ with eigenenergies $E_{\mu}$ and $E_{\nu}$, respectively, $\omega_{\mu \nu}$ is defined as $\omega_{\mu \nu} = (E_{\mu} - E_{\nu})/\hbar$. The Redfield tensor has the form~\cite{weiss2012quantum,blum2012density}
\begin{align}
 R_{\mu \nu \kappa \lambda} = \Gamma_{\lambda \nu \mu \kappa}^{+} + \Gamma_{\lambda \nu \mu \kappa}^{-} - \delta_{\nu \lambda} \sum \limits_{\alpha} \Gamma_{\mu \alpha \alpha \kappa}^{+} - \delta_{\mu \kappa} \sum \limits_{\alpha} \Gamma_{\lambda \alpha \alpha \nu}^{-} \label{eq:redfield_tensor}
\end{align}
with the rates given by~\cite{weiss2012quantum,blum2012density}
\begin{widetext}
\begin{align}
 & & \Gamma_{\lambda \nu \mu \kappa}^{+} = \hbar^{-2}\int \limits_{0}^{\infty} \mathrm{d}t \langle \bra{\lambda}H_{I, \text{SB}}(t)\ket{\nu} \bra{\mu}H_{I,\text{SB}}(0)\ket{\kappa} \rangle_{\text{bath}} {\rm e}^{- \im \omega_{\mu \kappa}t} \\
 & & \Gamma_{\lambda \nu \mu \kappa}^{-} = \hbar^{-2}\int \limits_{0}^{\infty} \mathrm{d}t \langle \bra{\lambda}H_{I,\text{SB}}(0)\ket{\nu} \bra{\mu}H_{I,\text{SB}}(t)\ket{\kappa} \rangle_{\text{bath}} {\rm e}^{- \im \omega_{\lambda \nu}t},
\end{align}
\end{widetext}
where $\langle\cdot\rangle_{\text{bath}}$ is the expectation value of the bath observable and $H_{I, \text{SB}}$ is the interaction Hamiltonian in the interaction picture with respect to the bath Hamiltonian,
\begin{align}
H_{I,\text{SB}}(t) & = ~ {\rm exp}\left\{ \im H_B t/\hbar \right\} H_{\text{SB}} ~ {\rm exp}\left\{ -\im H_B t/\hbar \right\} \\
& = ~ \sigma_x \sum \limits_q M_q {\rm e}^{\im q v t}\left( b_{-q}^{\dagger} {\rm e}^{\im \omega_q t} + b_q {\rm e}^{-\im \omega_qt} \right).
\end{align}
Using $\langle b_{-q}^{\dagger} b_l \rangle = \delta_{-q, l} n\left( \omega_q \right)$ with the Bose-Einstein distribution $n(\omega)$ the rates can be expressed as
\begin{widetext}
\begin{align}
 \Gamma_{\lambda \nu \mu \kappa}^{+} = M_{\lambda \nu \mu \kappa} \sum \limits_q \left| M_q \right|^2 \int \limits_{0}^{\infty} \mathrm{d}t \left\{ {\rm e}^{\im \left[ qv + \omega_q - \omega_{\mu \kappa} \right]t} n(\omega_q) + {\rm e}^{\im[qv - \omega_q -\omega_{\mu \kappa}]t} (n(\omega_q)+1) \right\} \\
 \Gamma_{\lambda \nu \mu \kappa}^{-} = M_{\lambda \nu \mu \kappa} \sum \limits_q \left| M_q \right|^2 \int \limits_{0}^{\infty} \mathrm{d}t \left\{ {\rm e}^{\im \left[ qv - \omega_q - \omega_{\lambda \nu} \right]t} n(\omega_q) + {\rm e}^{\im[qv + \omega_q -\omega_{\lambda \nu}]t} (n(\omega_q)+1) \right\},
\end{align} 
\end{widetext}
where $M_{\lambda \nu \mu\kappa} = \hbar^{-2} \braket{\lambda | \sigma_x | \nu} \braket{\mu| \sigma_x | \kappa}$. \\
For a subsonic motion, $v < c$, these rates are given by
\begin{widetext}
\begin{align}
  \Gamma_{\lambda \nu \mu \kappa}^{+} = \frac{\pi}{2} M_{\lambda \nu \mu \kappa} \left\{ \begin{array}{lr}  \frac{J\left(\omega_{\kappa \mu}^{-} \right)}{1-v/c} \left[n\left( \omega_{\kappa \mu}^{-} \right) + 1\right] + \frac{J\left( \omega_{\kappa \mu}^{+} \right)}{1+v/c} \left[n\left( \omega_{\kappa \mu}^{+} \right) + 1\right] &, \omega_{\kappa \mu} > 0 \\ 
 \frac{J\left( \omega_{\mu \kappa}^{+} \right)}{1+v/c} n\left( \omega_{\mu \kappa}^{+} \right)+ \frac{J\left( \omega_{\mu \kappa}^{-} \right)}{1-v/c} n\left( \omega_{\mu \kappa}^{-} \right) &, \omega_{\mu \kappa} > 0 \end{array} \right. \label{eq:sub_rates1}\\
 \Gamma_{\lambda \nu \mu \kappa}^{-} = \frac{\pi}{2} M_{\lambda \nu \mu \kappa} \left\{ \begin{array}{lr} \frac{J\left( \omega_{\lambda \nu}^{-} \right)}{1-v/c} \left[n\left( \omega_{\lambda \nu}^{-} \right) + 1\right] + \frac{J\left( \omega_{\lambda \nu}^{+} \right)}{1+v/c} \left[n\left( \omega_{\lambda \nu}^{+} \right) + 1\right] &, \omega_{\lambda \nu} > 0 \\ 
 \frac{J\left( \omega_{\nu \lambda}^{+} \right)}{1+v/c} n\left( \omega_{\nu \lambda}^{+} \right)+ \frac{J\left( \omega_{\nu \lambda}^{-} \right)}{1-v/c} n\left( \omega_{\nu \lambda}^{-} \right) &, \omega_{\nu \lambda} > 0 \end{array} \right., \label{eq:sub_rates2}
\end{align}
\end{widetext}
with $\omega_{\alpha \beta}^{\pm} = \omega_{\alpha \beta}/(1\pm v/c)$ and where $J(\omega) = \sum \limits_q \left| M_q\right|^2 \delta(\omega - \omega_q)$ denotes the spectral density and it was used that $\int_0^{\infty}{\rm e}^{\im \omega t} \mathrm{d}t = \pi \delta(\omega)$, where the imaginary parts resulting from principal value integrals are neglected as they manifest themselves as Lamb shifts.

For a supersonic motion, $v > c$, and a vanishing temperature of the phonon bath, the rates are given by
\begin{align}
 & & \Gamma_{\lambda \nu \mu \kappa}^{+} = \frac{\pi}{2} M_{\lambda \nu \mu \kappa} \left\{ \begin{array}{lr} \frac{J\left( \omega_{\kappa \mu}^{+} \right)}{v/c+1} &, \omega_{\kappa \mu} > 0 \\ \frac{J\left( \omega_{\mu \kappa}^{-} \right)}{v/c - 1} &, \omega_{\mu \kappa} > 0 \end{array} \right. \label{eq:rates1}\\
 & & \Gamma_{\lambda \nu \mu \kappa}^{-} = \frac{\pi}{2} M_{\lambda \nu \mu \kappa} \left\{ \begin{array}{lr} \frac{J\left( \omega_{\lambda \nu}^{+} \right)}{v/c+1} &, \omega_{\lambda \nu} > 0 \\ \frac{J\left( \omega_{\nu \lambda}^{-} \right)}{v/c - 1} &, \omega_{\nu \lambda} > 0 \end{array} \right. . \label{eq:rates2}
\end{align} 

From Eqs~\ref{eq:sub_rates1} and~\ref{eq:sub_rates2} and Eqs~\ref{eq:rates1} and~\ref{eq:rates2}, respectively, the Redfield tensor Eq~\eqref{eq:redfield_tensor} can be computed for each corresponding case. 

For the steady state solution of the Bloch-Redfield Eqs~\ref{eq:BRE} one finds vanishing off-diagonal matrix elements, $\rho_{10} = \rho_{01} = 0$, in both sub- and supersonic cases, as expected due to spin decoherence. The diagonal matrix elements can be expressed as
\begin{align}
 \frac{\rho_{11}}{\rho_{00}} = \frac{\Gamma_{12}}{\Gamma_{21}} \label{eq:steady_state_gen}
\end{align}
with the absorption rate $\Gamma_{12} = R_{1111} = \Gamma_{1221}^{+} + \Gamma_{1221}^{-}$ and the emission rate $\Gamma_{21} = R_{1122} = \Gamma_{2112}^{+} + \Gamma_{2112}^{-}$. 

For subsonic motion, $v < c$, these rates explicitly read

\begin{widetext}
\begin{align}
\Gamma_{12} &= \frac{\pi}{2}M_{1221}\left\{ \frac{J\left( \omega_{21}^{+}\right)}{1+v/c}n\left( \omega_{21}^{+}\right) + \frac{J\left( \omega_{21}^{-} \right)}{1-v/c}n\left( \omega_{21}^{-} \right) \right \}  \\ \Gamma_{21} &= \frac{\pi}{2}M_{1221} \left\{ \frac{J\left( \omega_{21}^{+}\right)}{1+v/c}\left[ n\left( \omega_{21}^{+} \right) + 1 \right] + \frac{J\left( \omega_{21}^{-} \right)}{1-v/c}\left[ n\left( \omega_{21}^{-}\right) + 1 \right] \right\}. 
\end{align}
\end{widetext}

The steady state Eq~\eqref{eq:steady_state_gen} is clearly non-thermal in the sense that the effective temperature $T'$ of the DQD, given via
\begin{align}
 \frac{\rho_{11}}{\rho_{00}}=\exp\left(- \frac{\hbar \omega_{10}}{k_B T'} \right) \label{eq:eff_temp}, 
 %={\rm e}^{-\hbar \omega_{10}/(k_B T')}={\rm e}^{-\frac{\hbar \omega_{10}}{k_BT'}} 
\end{align}
does not in general coincide with the temperature of the phonon bath the DQD is in contact with, $T \neq T'$. In this scenario the rates and thus the non-thermality can be easily explained via the Doppler-effect: Phonons participating in transitions in the DQD have a frequency in the rage $\Omega = [\omega_{10} - \Delta \omega/2, \omega_{10}+\Delta \omega/2]$ with the line width $\Delta \omega$. As these frequencies are measured in the reference frame of the DQD, and the speed of sound $c$ and the speed of the DQD add up/subtract leading to a Doppler-shift $\omega \to (1\pm v/c) \omega$, there are in fact two frequency ranges involved in transitions, namely $\Omega/(1\pm v/c)$ measured in the bath frame. Each of them contribute to the absorption and emission rate in the usual manner, where the prefactors $(1\pm v/c)^{-1}$ are due to the Doppler-shifted line width. Furthermore, for a zero temperature phonon bath the absorption rate also vanishes, $\Gamma_{12} = 0$, since there are no phonons present which could excite the DQD and thus the DQD equilibrates with the phonon bath by relaxing to its ground state and thus having vanishing effective temperature.

In the case of supersonic motion, $v > c$, where we restricted the analyses to a vanishing bath temperature $T = 0$, the absorption and emission rates are now given by
\begin{align}
 & & \Gamma_{12} = \pi \frac{M_{1221}}{v/c - 1} J\left( \omega_{21}^{-} \right) \\
 & & \Gamma_{21} = \pi \frac{M_{1221}}{v/c + 1}J\left( \omega_{21}^{+} \right).
\end{align}
Even though the bath is at zero temperature, the effective temperature of the DQD given via Eq~\eqref{eq:eff_temp} is non-vanishing, because the absorption rate does not vanish. We argue that this motion-enhanced character of the density matrix can be attributed to the presence of excess particles in form of piezophonons originating from particle creation, which excite the DQD and lead it to a steady state which is not in thermal equilibrium with the bath.

The use of Stern-Gerlach gates or other spin-to-charge conversion methods allows for the readout of the electron spin from which the effective DQD temperature can be computed. Equilibration is expected to be achieved after a time $\Gamma_{12} + \Gamma_{21}$ (see, e.g.~\cite{breuer2002theory}), where the $\Gamma$-terms denote the rate for a spin-flip with emission and absorption of a piezophonon, respectively. An upper bound for the equilibration time is given by  the rate for spontaneous emission of a piezophonon. A rough estimation of this rate for the present set-up is given by Eq (8) of Ref~\cite{khaetskii2001spin}, but with an additional factor $\exp\left\{ - 2d (g\mu_B B)/(\hbar c) \right\}$, where $d$ denotes the distance between the two substrates, taking into account the exponential decay of the piezoelectric field accompanying the piezophonons~\cite{aizin1998screening, simon1996coupling}, and the energy conservation, $\hbar c k = g \mu_B B$~\cite{Note4}. For a magnetic field $B = 1$ T, the equilibration rate is of the order of $1 \text{ s}^{-1}$. The steady state of the electron spin could be achieved while the electrons are stored in a storage ring as it is shown in Fig~\ref{fig:setup}. Efficient electron transport over macroscopic distances has been shown in ~\cite{Note5}: The lengths $l, L$ of the storage ring can be chosen arbitrarily in the sub-mm regime. Future conveyor belts for electrons using serpentine-shaped SAW waveguides~\cite{boucher2014ring, adkins1971long} could provide an alternate route to reach equilibration. In comparison, however, the present set-up details make the proposed storage ring more feasible.

We note that the use of quantum field theory for the SAW field whose propagation is described by the metrics with line elements given in Eq~\eqref{eq:static_metric} and Eq~\eqref{eq:dynamic_metric}, respectively, is a more natural route to study particle creation in this system. A Bogoliubov transformation can be computed from the two sets of positive and negative frequency modes used for the reference frame of the substrate and the moving observer, respectively.
Particle creation occurs if positive (or negative) frequency modes from one set appears as mixture of both positive and negative frequency modes from the other set. We do not use this approach as it is out of scope of the present study and leave it for future work.

\section{Conclusion}
\label{sec:conc}
We have presented a new semiconductor analogue system to simulate quantum effects in general relativity. Particle creation in this system is expected for a DQD moving with constant speed in the measurement substrate. The detailed steps to achieve this are discussed, including the charge density modulation of a two dimensional electron gas which is responsible for the change in speed of sound for SAWs travelling on the substrate. This in turn leads to \ahorizon~for SAWs seen by the DQD. 

We analyzed a measurement scheme to detect the created particles using the DQDs, whose steady-state spin populations differ from that of a thermal state due to their interaction with the created piezophonons. We stress that a number of different alternatives to observe the evanescent waves could be pursued, as their detection is well developed in fields such as biosensing~\cite{Sapsford2008}. However, these schemes must be adapted to allow for the characterization of the quantum nature of the associated SAW phonons. 

We note that the origin of the excess phonons in the moving frame is not clear. Further work to determine the cosmological nature of this radiation is still required.

\section{Acknowledgements}
We thank Malte Henkel, Rodrigo Turcati, David Edward Bruschi and specially Germain Rousseaux for helpful discussions. The authors would like to acknowledge the reviewers for pointing out important issues that lead to significant improvement of this work. We acknowledge partial support by the European Union through project ScaleQIT. B.G.T. also acknowledges support from Fapesc and CNPq INCT-IQ (465469/2014-0).

%%%%%%%%%%%%%%%%%%%%%%%%%%%%%%%%%%%%%%%%%%%%%%%%%%%%%%%%%%%%%%%
%%%%%%%%%%%%%%%%%%%%%%%%%%%%%%%%%%%%%%%%%%%%%%%%%%%%%%%%%%%%%%%
%%%%%%%%%%%%%%%%%%%%%%%%%%%%%%%%%%%%%%%%%%%%%%%%%%%%%%%%%%%%%%%
%%%%%%%%%%%%%%%%%%%%%%%%%%%%%%%%%%%%%%%%%%%%%%%%%%%%%%%%%%%%%%% 
\appendix

\section{Approximation of the SAW Speed}
\label{app:SAW_speed}
For convenience, the 2DEG density is denoted by $n$ in the following. In principle, an expression for the SAW speed $c(x)$ in terms of $n$ can be obtained from the implicit equation Eq~\eqref{eq:SAW_speed_1}. I.e. $c(x) = f(n, \partial_xn)$ with a particular function $f$, and derivation of this equation with respect to $x$ gives
\begin{align}
 \pd{c}{x} = \pd{f}{n} \pd{n}{x} + \pd{f}{\left(\partial_xn\right)}\spd{n}{x}{x}.
 \label{eq:c_prime}
\end{align}
Expanding $n(x)$ (see Eq~\eqref{eq:2DEG_dens}) in a Taylor series around the gate's edge $x = 0$ up to third order and solving for the roots, one obtains a value of approximately $4 \kappa_s^{-1}$ for the size of the region, where $n(x)$ changes from 0 to its maximum value. Outside this region $\partial_xn$ and $\partial^2_x n$ vanish. According to Eq~\eqref{eq:c_prime}, the SAW speed consequently only changes around the gate's edge, too. Thus, as already mentioned, a good approximation for $c(x)$ is a piecewise constant behavior in the regions $|x| > 2\kappa_s^{-1}$, and a linear adjustment in between $|x| < 2\kappa_s^{-1}$. Deep inside the gate, i.e. for $x < 2\kappa_s^{-1}$, the corresponding value for the SAW speed is obtained by letting $n \to \infty$ and $\partial_x n \to 0$ in Eq~\eqref{eq:eff_perm}, giving $d_{\text{eff}} = d$ (see Eq~\eqref{eq:eff_d}) and consequently $c = c_0 = \sqrt{d/\rho}$ (see Eq~\eqref{eq:SAW_speed_1}). In the region far away from the gate, i.e. $x > 2\kappa_s^{-1}$, the 2DEG density is vanishing $n = \partial_xn = 0$, and, with Eqs~(\eqref{eq:eff_perm}-\eqref{eq:SAW_speed_1}), give $c = c_0 \sqrt{1 + K^2} \approx c_0 \left( 1 + \tfrac{1}{2}K^2\right)$, where a small piezoelectric coupling constant $K^2 = e^2/(\varepsilon d) \ll 1$ is assumed, which is valid for nearly all piezoelectric materials~\cite{wixforth1989surface}. 

%\bibliographystyle{apsrev4-1}
%\bibliography{hawkingReDraft_bib}

\begin{thebibliography}{10}
\bibitem{note6}
In a flat space-time particle creation does not occur for two inertial observers because of the Lorentz invariance. In analogue models, however, the existence of a preferred speed (i.e. speed of sound in the present set-up) breaks Lorentz invariance. The motion of the moving observer can be regarded as an accelerated motion while passing the region where the SAW speed changes locally.
\end{thebibliography}

\end{document}